\documentclass[twocolumn,preprint]{revtex4-1}
\usepackage{graphicx}
\usepackage{pstricks}
\usepackage{amsmath}
\usepackage{bm}

\usepackage[position=top]{subfig}
\usepackage{setspace}
\usepackage{floatrow}
\usepackage{epstopdf}

\begin{document}
\title{Pentaoctite phase: A new group V allotrope}

\author{A. L. Rosa}
\affiliation{Federal University of Goi\'as, Institute of Physics, Campus Samambaia, 74690-900, Goi\^ania, Goi\'as, Brazil}
\affiliation{Bremen Center for Computational Materials Science, University of Bremen, Am Fallturm 1, 28359 Bremen, Germany}
\email{andreialuisa@ufg.br}
\author{E. N. Lima}
\affiliation{Federal University of Rondon\'opolis,
Av. dos Estudantes, 5055,  Cidade Universit\'aria, 78736-900 Rondon\'opolis, Mato Grosso, Brazil}
\email{erika@ufr.edu.br}
\author{Th. Frauenheim}
\affiliation{Bremen Center for Computational Materials Science, University of Br
emen, Am Fallturm 1, 28359 Bremen, Germany}
\affiliation{Shenzhen Computational Science and Applied Research Institute, Shen
zhen, China}
\affiliation{Beijing Computational Science Research Center, Beijing, China}

\begin{abstract}

  By performing firt-principles electronic calculations we propose new
  a phase of group-V allotropes of antimonene, arsenene and phosphorene
  in the pentaoctite structure. By calculating the phonon spectra, we
  show that all these phases are stable.  Whereas these structures
  have a indirect band gap, they can be made direct gap materials by
  applying external strain. GW calculations of the dielectric function
  demonstrate that all these structures have an absorption spectrum in
  the visible region, which could be useful for
  group-V optoelectronics.

\end{abstract}

%\pacs{PACS numbers: 75.60.Ej, 64.60.Ak}

\maketitle

\section{Introduction}

Since the discovery of graphene\,\cite{Geim:07,Novoselov:04} much
attention has been devoted to discover new two-dimensional materials
due to their exceptional properties such as high electrical
conductivity, and mechanical robustess. However, opening a band gap in
graphene has been proofed rather difficult, thus limiting its
applications in electronic devices. Therefore, topological defects
have been often used to tune the electronic properties of
two-dimensional carbon materials. In particular, structural
pentagonal, heptagonal and octagonal rings have been considered as
possible defects in carbon
nanostructures\,\cite{NunesPRB2010,NunesNanotech2012,NunesNL2012,KawazoePNAS2015,SantosNat2020}
Indeed defects composed of pentagons and
octagons embedded in a perfect graphene have been observed in
graphene\,\cite{PRB2014,NatureNano2010}.  The advantage of these lower symmetry
structures is that they could be more
easily functionalized than graphene and therefore are
promising for applications in optoelectronics\,\cite{KawazoePNAS2015}.

Recently, the existence of an allotrope phase of bismuthene called
pentaoctite, in which all hexagonal rings are replaced by either
pentagons or octagons has been
proposed\,\cite{LimaJPCM2019,LimaNL2016}.  These structures show a
sizebale band gap, can be stable under strain and have topological insulator behavior with protected
surface non-trivial Dirac states.

In this work we extend our investigations of this allotrope phase to
phosphorene\,\cite{BlueP2017},
arsenene\,\cite{SciRep2019,AsSciRep2016} and
antimonene\,\cite{Sb2013,NL2019}. Our first-principles calculations show that these
two-dimensional structures are metastable against their respective hexagonal
phases, but have relatively low formation energies. In particular
group-V pentaotite can become a direct gap materials under tensile or
compressive strain. Our calculated dielectric function show that all
structures have absorption edges in the visible region, making these
materials suitable for optoelectronic applications.

\section{Methodology}

In our calculations, the first principles geometry optimizations and
the electronic structure calculations were performed using the density
functional theory (DFT) \cite{Hohenberg:64,Kohn:65}, as implemented in
the Vienna \textit{ab initio} simulation package VASP \cite{VASP}. The
generalized gradient approximation (GGA)~\cite{GGA} is employed to
describe the exchange and correlation potential. The interactions
between the valence electrons and the ionic cores are treated within
the projector augmented wave (PAW)~\cite{PAW1,PAW} method. The
electronic wave functions are expanded on a plane-wave basis with an
energy cutoff of 500 eV. The Brillouin zone (BZ) is sampled using a
(10$\times$10$\times$1) $\Gamma$-centered Monkhorst-Pack\,\cite{mk}
grid. The system is modelled using supercells repeated periodically
along the nanostructure plane with a vacuum region of 12 {\AA} perperdicular to the surface plane in order to avoid the interactions between periodic images of the supercell. Spin-orbit coupling was included in
all electronic structure calculations. Phonon properties were carried
out employing the density functional perturbation theory (DFPT)
method, as implemented in the PHONOPY code\,\cite{phonopy}. The
calculation of the dielectric function was performed using the GW
method\,\cite{Shishkin:07} with a (6$\times$6$\times$1) {\bf k}-points
mesh and energy cutoff of 400 eV. 

\section{Results and Discussions}

\begin{table*}[ht]
\caption{GGA-PBE optimized structure parameters of pentaoctite P, As, and Sb. $a$ and $b$ are the lattice constants and $h$ is the buckling height. The bang gaps E$_{\rm g}$ are given in eV.  Z$_2$ is the topological invariant.}
\begin{tabular*}{16cm}{@{\extracolsep{\fill}}lccccccc}
\hline
            & $a$ (\AA) & $b$(\AA) & $h$(\AA) & \multicolumn{3}{c}{E$_{\rm g}$} & Z$_2$\\ [0.5ex]
            &           &          &          &  PBE & HSE & G$_0$W         &        \\
\hline 
P           & 9.00      & 6.33     & 1.41     & 1.01 & 1.34 & 2.94 & 0\\ 
As          & 9.76      & 7.05     & 1.63     & 1.13 & 1.36 & 2.37 & 0\\
Sb          & 11.07     & 8.00     & 1.92     & 0.83 & 1.01 & 2.60 & 0\\ 
\hline 
\end{tabular*}
\label{table:parameters} 
\end{table*}

\begin{figure}[ht!]
\begin{center}
 \includegraphics[width = 8cm,scale=1, clip = true]{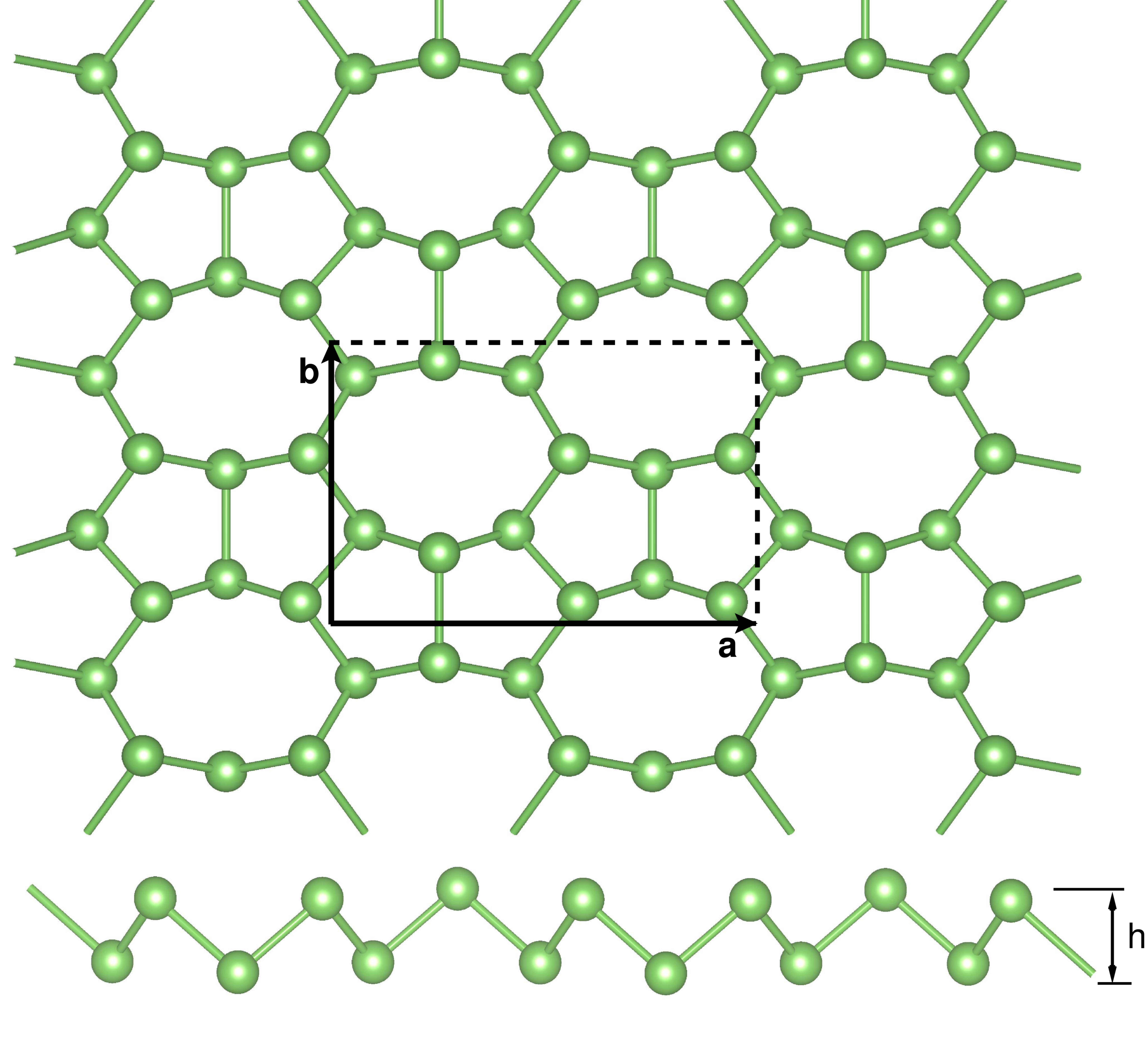}
\end{center}
\caption{\label{fig:estrutura-pentaoctite} Top and side views of relaxed pentaoctite structure. The unit cell with the lattice vectors and buckling $h$ (lower panel) are shown.}
\end{figure}

The crystal structure of a typical pentaoctite sheet is shown in Fig.
\,\ref{fig:estrutura-pentaoctite}. The top view shows that the new
phase is composed of two side-sharing pentagons connected to octagon
rings. The unit cell contains 12 atoms. The side view (lower panel)
shows the buckling height (h). Typically, the nearest-neighboring atoms
belong to different sublayers. The equilibrium in-plane parameters of
pentaoctite sheet of P, As, and Sb are $a$=9.00\,{\AA},  $b$=6.33 \AA;
$a$=9.76 {\AA}, $b$=7.05 {\AA} and $a$= 11.07 {\AA}, $b$= 8.00
{\AA}, respectively.  Buckling increases with atomic number and is
larger for Sb. These lattice parameters increase as the atomic size
increases.  The optimized structural parameters are shown in Table
\ref{table:parameters}.

\floatsetup[figure]{style=plain,subcapbesideposition=top}

\begin{figure*}[ht!]
\begin{center}
\sidesubfloat[]{\includegraphics[width=5cm,scale=1.0,clip, keepaspectratio]{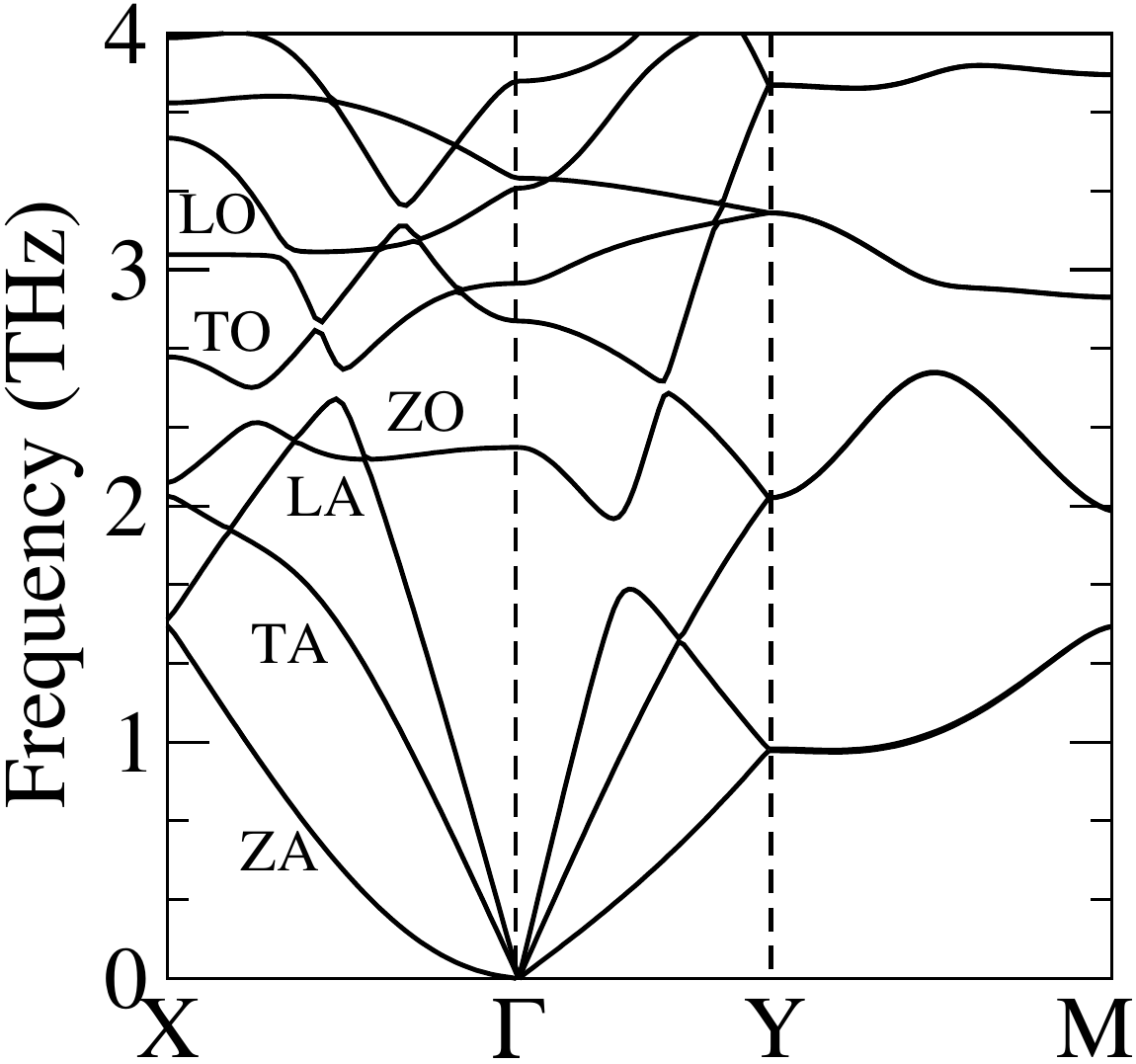}}
\sidesubfloat[]{\includegraphics[width=5cm,scale=1.0,clip, keepaspectratio]{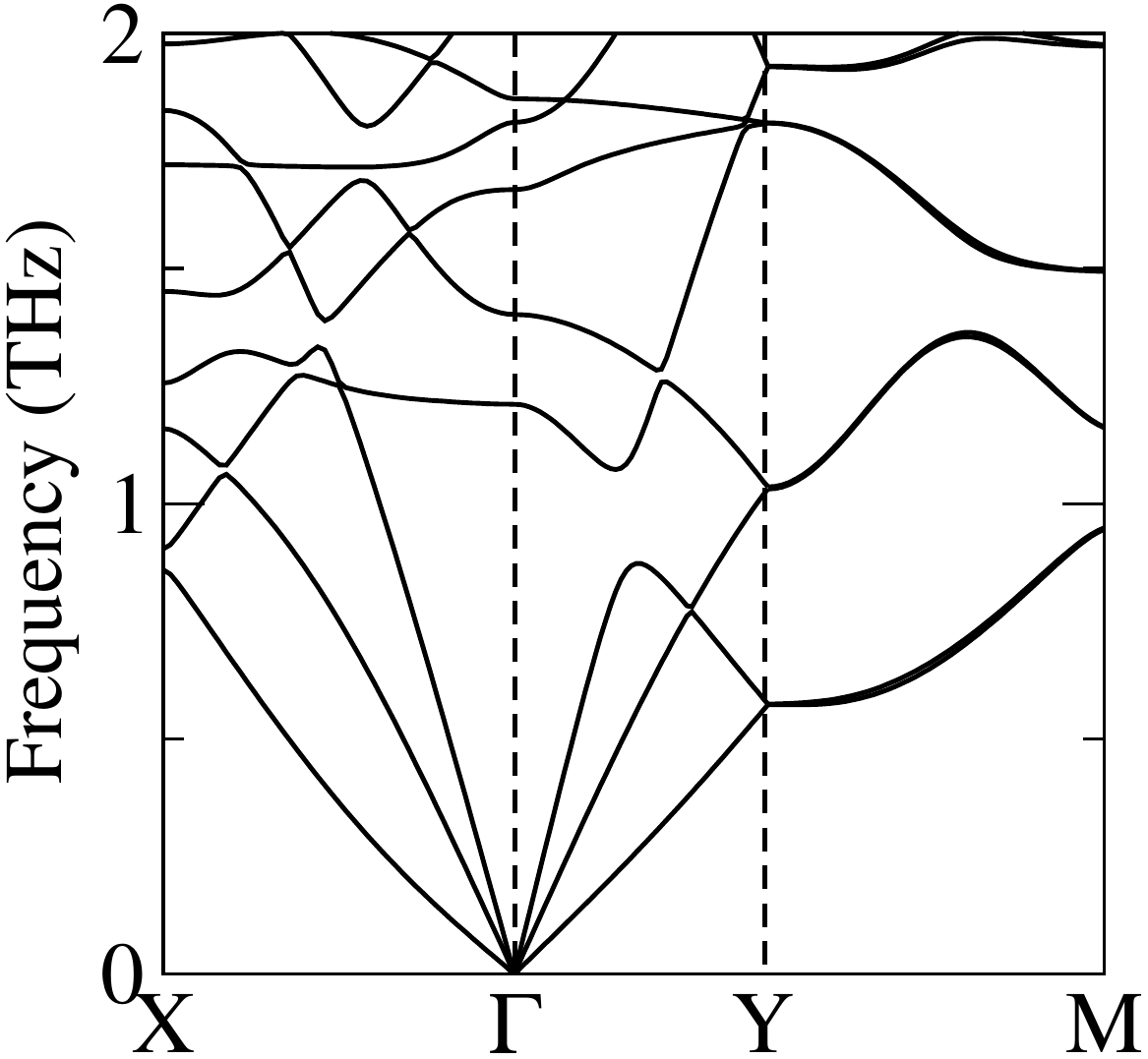}}
\sidesubfloat[]{\includegraphics[width=5cm,scale=1.0,clip, keepaspectratio]{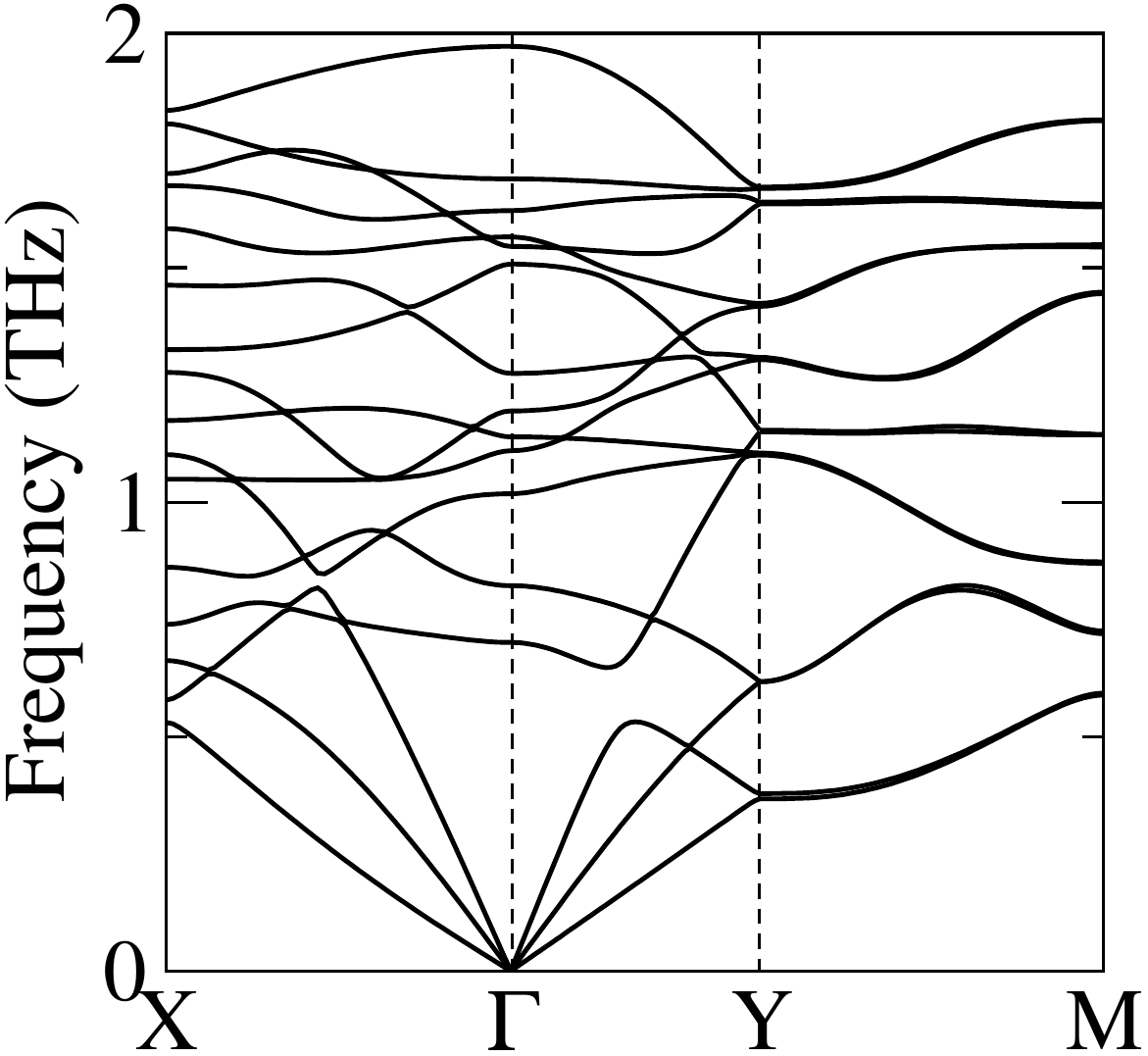}}
\end{center}
\caption{\label{fig:phonons}Phonon dispersion relation for pentaoctite (a) P, (b) As, and (c) Sb.} 
\end{figure*}

In order to study the stability of pentaoctite sheets of P, As, and
Sb, we examined the phonon dispersions along the
high-symmetry lines in the Brillouin zone (BZ) (Fig.
\ref{fig:phonons}). No imaginary modes are present, confirming that
the freestanding pentaoctite sheet is dynamically stable. Similarly to
other 2D layered materials\,\cite{WANG201677,Zhang2372}, there are
three distinct acoustic modes in the phonons spectra of
pentactite. The in-plane longitudinal acoustic (LA) and transverse
acoustic branches (TA) have linear dispersions near the $\Gamma$
point, whereas the out-of-plane acoustical modes (ZA) has quadratic
dispersion. The quadratic ZA mode in the long-wavelength region is
closely associated with the bending rigidity and lattice heat capacity
of the nanosheets, as discussed in
pentagraphene\,\cite{KawazoePNAS2015} Also, there is a hybridization
of ZA and ZO modes in all phonon dispersions. As a matter of
comparison, in hexagonal As, the optic and acoustic branches are well
separated by a gap\,\cite{SciRep2019}. Although the presence of
imaginary frequencies is not seen, the parabolic behavior of one of
the acustic modes close to the $\Gamma$-point is present
Fig.\,\ref{fig:phonons}. In general, the ZA mode is very soft in 2D
materials, and a slight reduction of the lattice constant may result
in imaginary phonon frequencies near the $\Gamma$ point. In particular
for pentaoctite phophosrene this
behavior is more emphasized, as shown in Fig.\,\ref{fig:phonons} (c).

Additionally we have calculated the energy formation of pentaoctite
sheets, $\Delta$E, with respect to the honeycomb structures. The
formation energy is defined as ${\rm \Delta E = E_{total} -
  N_{atom}\times\mu_{atom}/N_{atom}}$,where E$_{total}$ is the total
energy of pentactite (P, As, and Sb), N$_{atom}$ is the total number of
atoms in the crystal structure, and $\mu_{atom}$ is the energy per
atom calculated for hexagonal honeycomb structure. The relative
formation energies of pentaoctite sheets of P, As, and Sb are 46
meV/atom, 49 meV/atom, and 62 meV/atom, respectively.

\begin{figure}[ht!]
  \centering
   \sidesubfloat[]{\includegraphics[width=5cm,scale=1.0,clip, keepaspectratio]{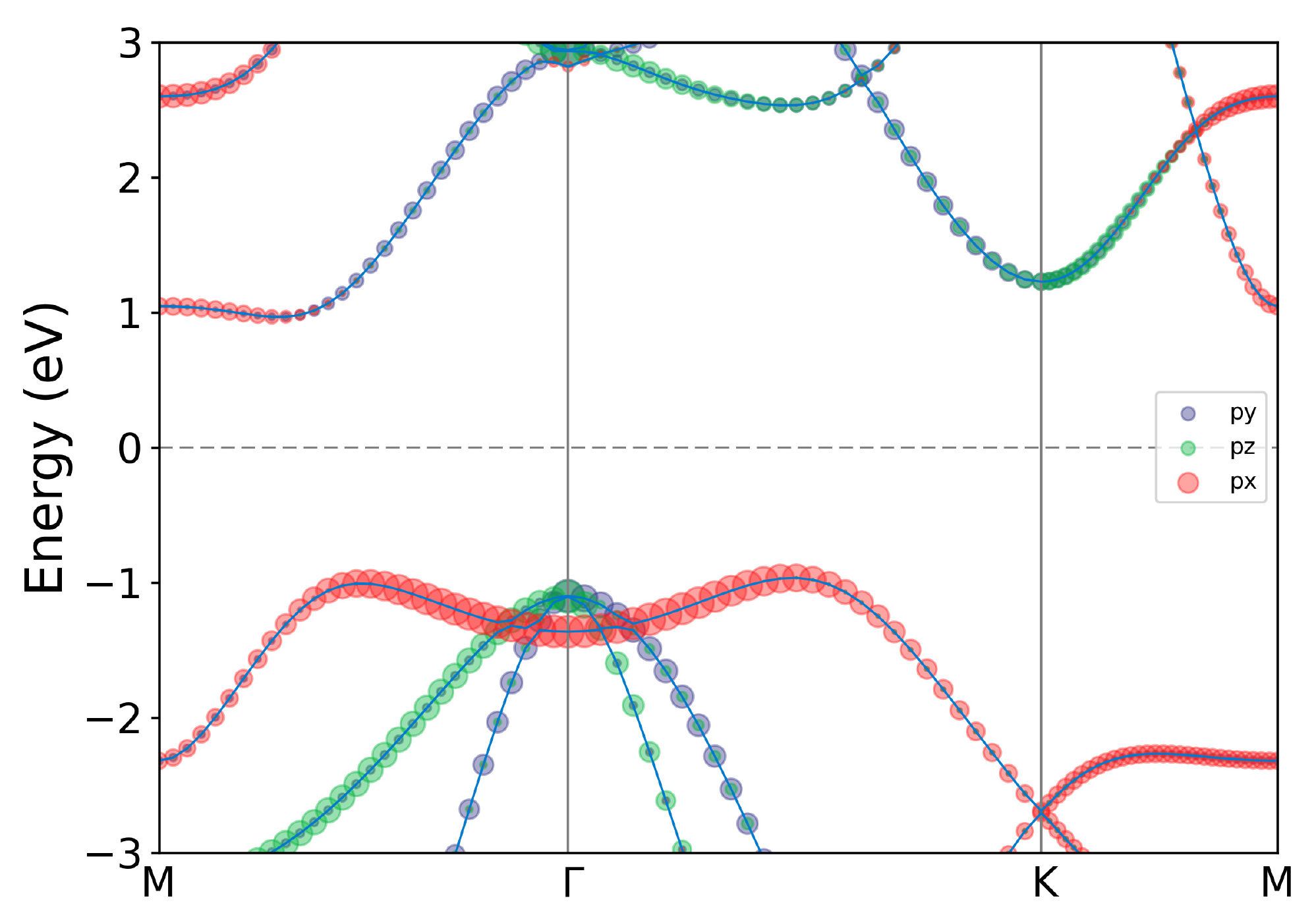}}\\
\sidesubfloat[]{\includegraphics[width=5cm,scale=1.0,clip, keepaspectratio]{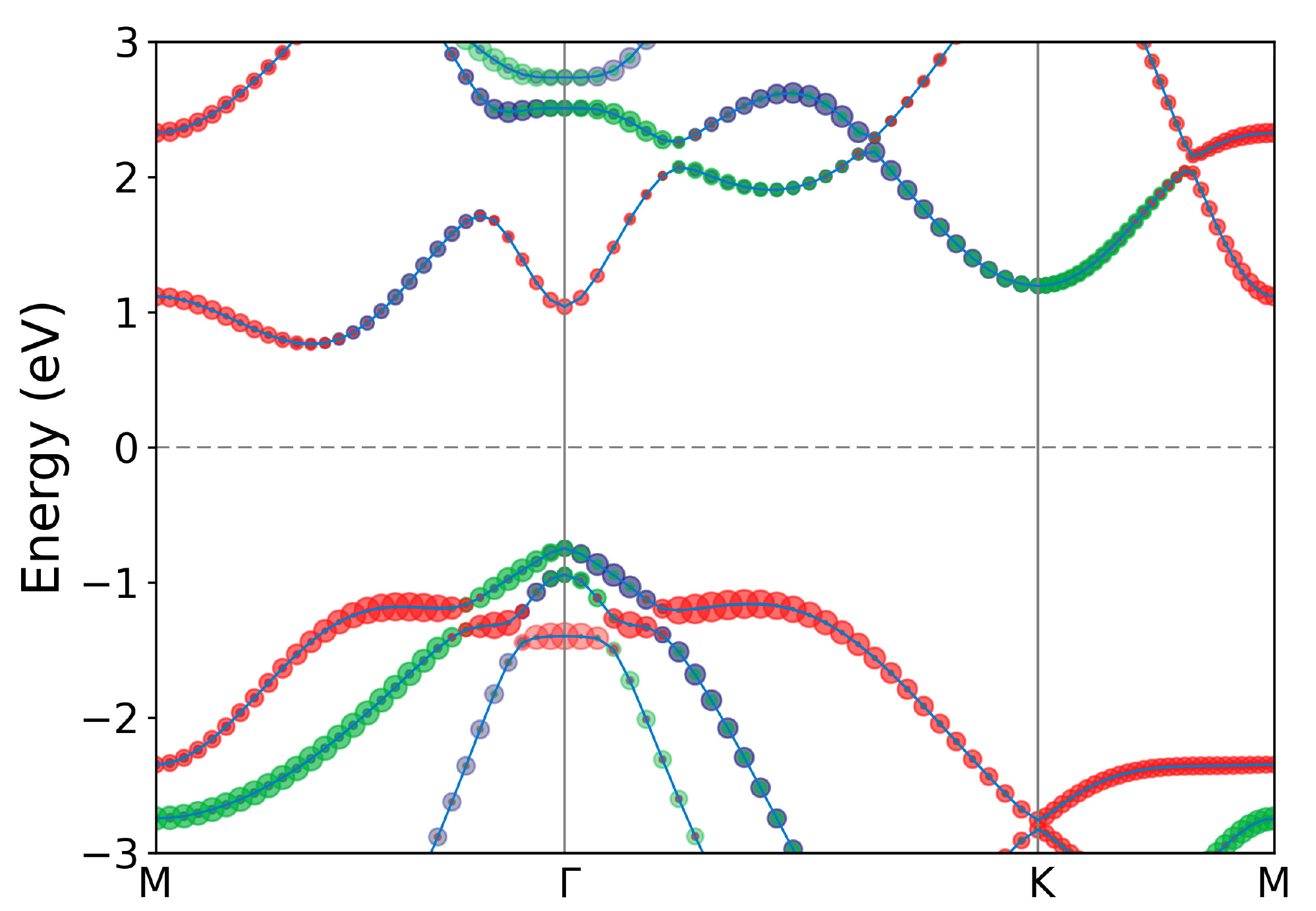}}\\
\sidesubfloat[]{\includegraphics[width=5cm,scale=1.0,clip, keepaspectratio]{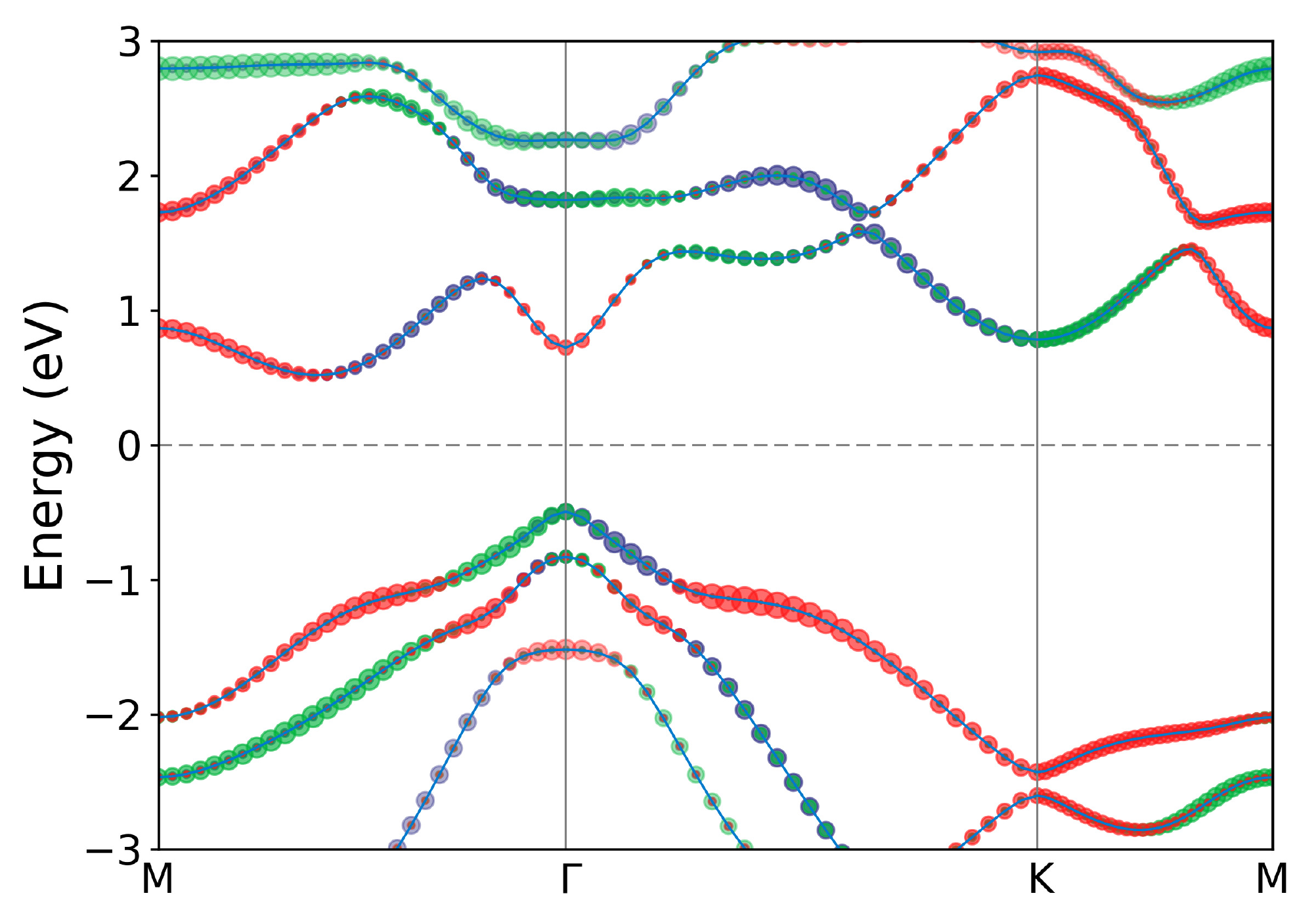}}\\
\caption{\label{fig:bandas_hexagonal}
Electronic band structure, with different orbital contributions of hexagonal (a) P, (b) As, and (c) Sb and pentaoctite (d) P, (e) As and (f) Sb at zero strain.  Red, green, and blue colors represent contribution from p$_x$, p$_y$, and p$_z$ orbitals, respectively.} 
\end{figure}

In order to compare the band structures of hexagonal and pentaoctite
structures, we have calculated the orbital projected band structure of
hexagonal blue phosphorene, arsenene and atimonene. These are shown in
Fig.\,\ref{fig:bandas_hexagonal}. Hexagonal blue phosphorene is a
indirect band gap semiconductor with the conduction band minimum (CBM)
located between the $\Gamma$ and M points while valence band maximum
(VBM) is located between the K and $\Gamma$ points, as seen in
Fig.\,\ref{fig:bandas_hexagonal}(a). Most contribution at VBM comes
from p$_x$ orbitals. Band gap is around 1.91 eV, in
good agreement with previous GGA
investigations\,\cite{Ospina2016,SciRep2016}. The band gap of
pentaoctite-P is around 1.01 eV (1.34 eV with HSE). These values
are reported in Table\,\ref{table:parameters}. Buckled hexagonal arsenene is an indirect bandgap semiconductor with
the VBM at the $\Gamma$-point and the CBM along the
$\Gamma$-M-direction, as shown in
Fig.\,\ref{fig:bandas_hexagonal}(b). The calculated indirect band gap
is 1.53 eV in agreement with previous calculations\,\cite{PCCP2018}.
HSE indirect has been reported to be
2.89\,\cite{SciRep2019}. Furthermore a free standing monolayer of
hexagonal As has negative frequencies and therefore though the
structure is energetically stable, it is dynamically not
stable\,\,\cite{SciRep2019,PRB2015}. On the other hand, our results
show that the calculated band gaps of pentaoctite-As are 1.13\,eV with
GGA and 1.36\,eV with HSE, as shown in Table\,\ref{table:parameters}.

\begin{figure*}[ht!]
  \centering
\sidesubfloat[]{\includegraphics[width=4cm,scale=1.0,clip, keepaspectratio]{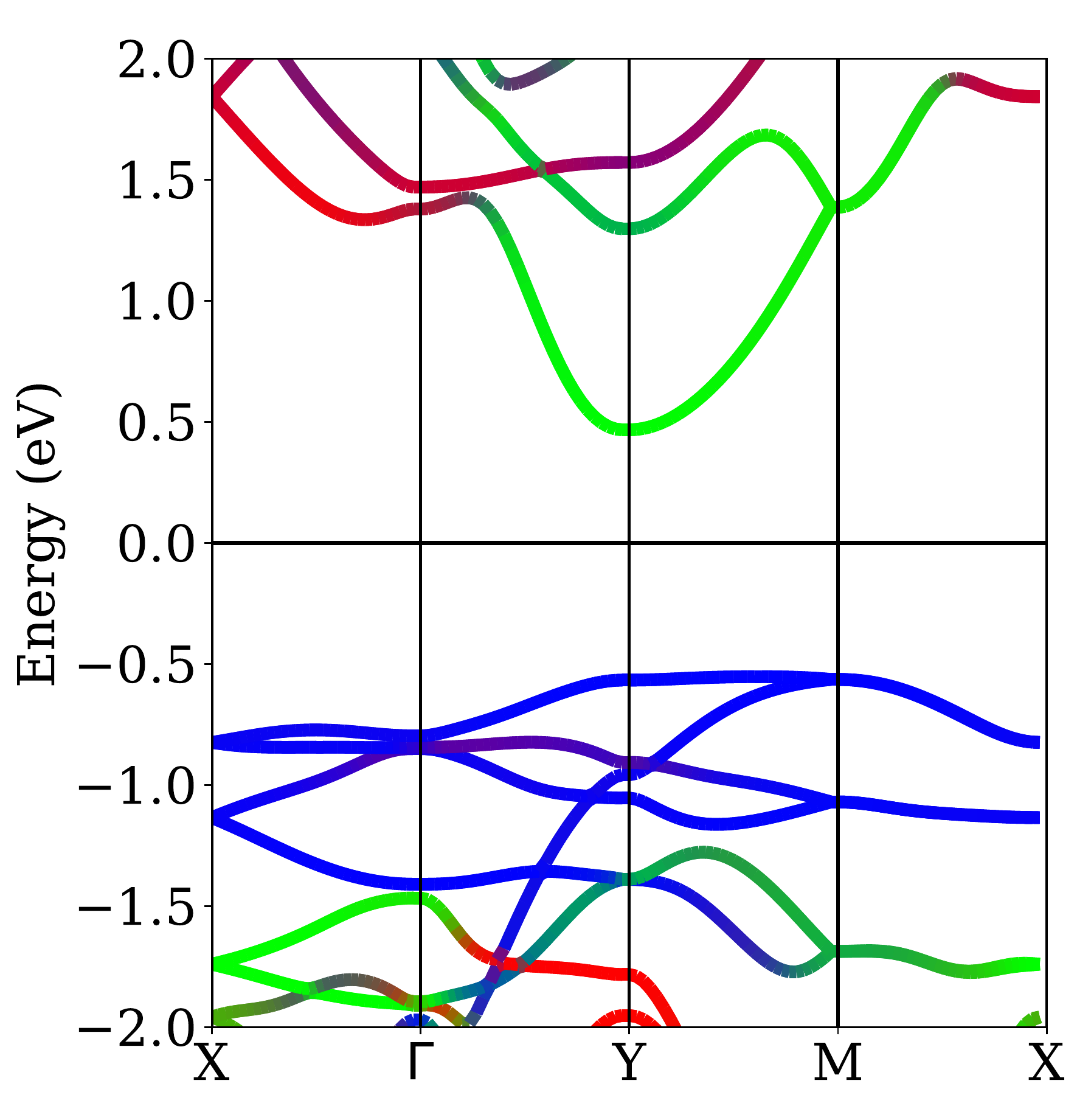}}\hfill
\sidesubfloat[]{\includegraphics[width=4cm,scale=1.0,clip, keepaspectratio]{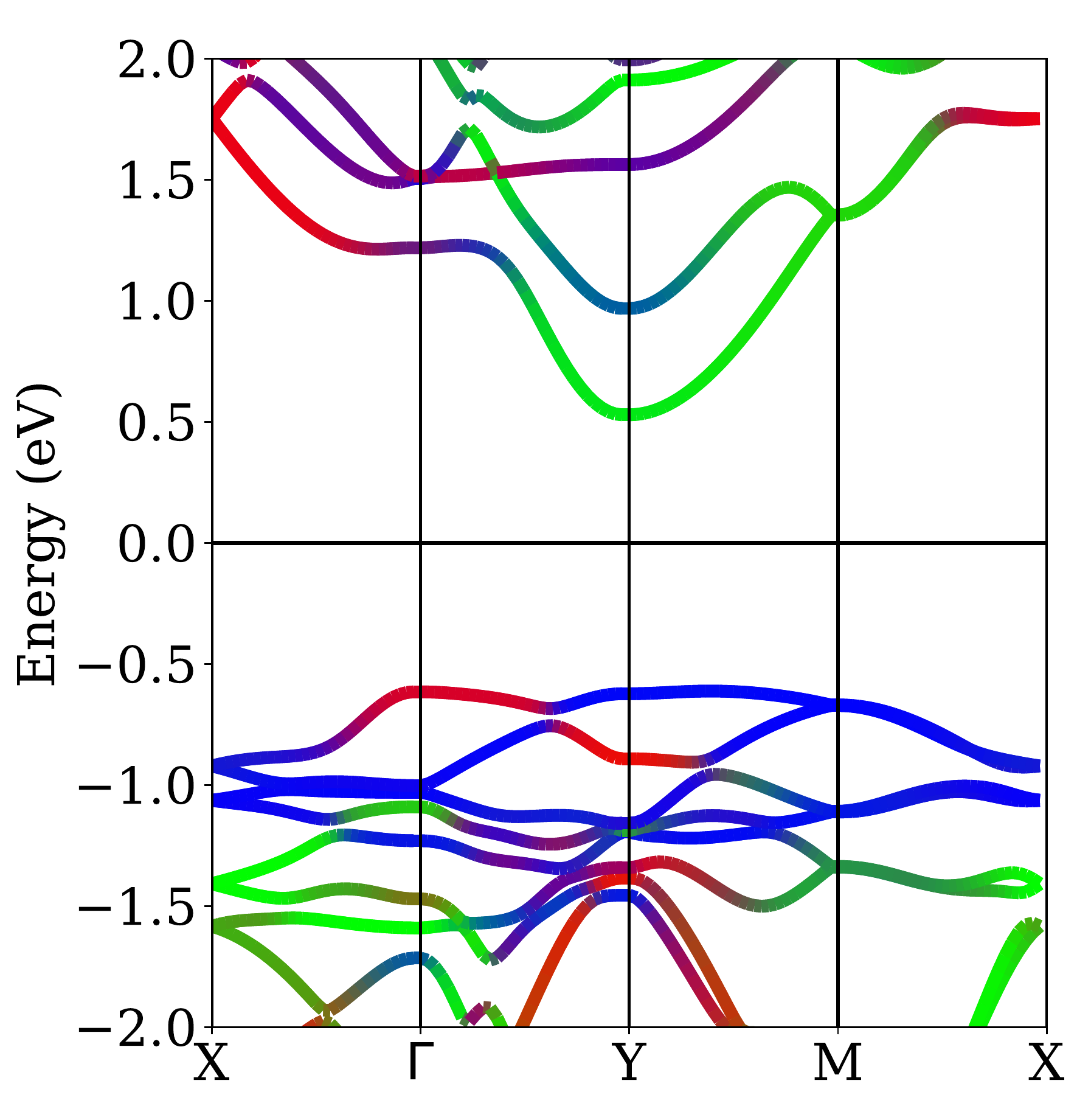}}\hfill
\sidesubfloat[]{\includegraphics[width=4cm,scale=1.0,clip, keepaspectratio]{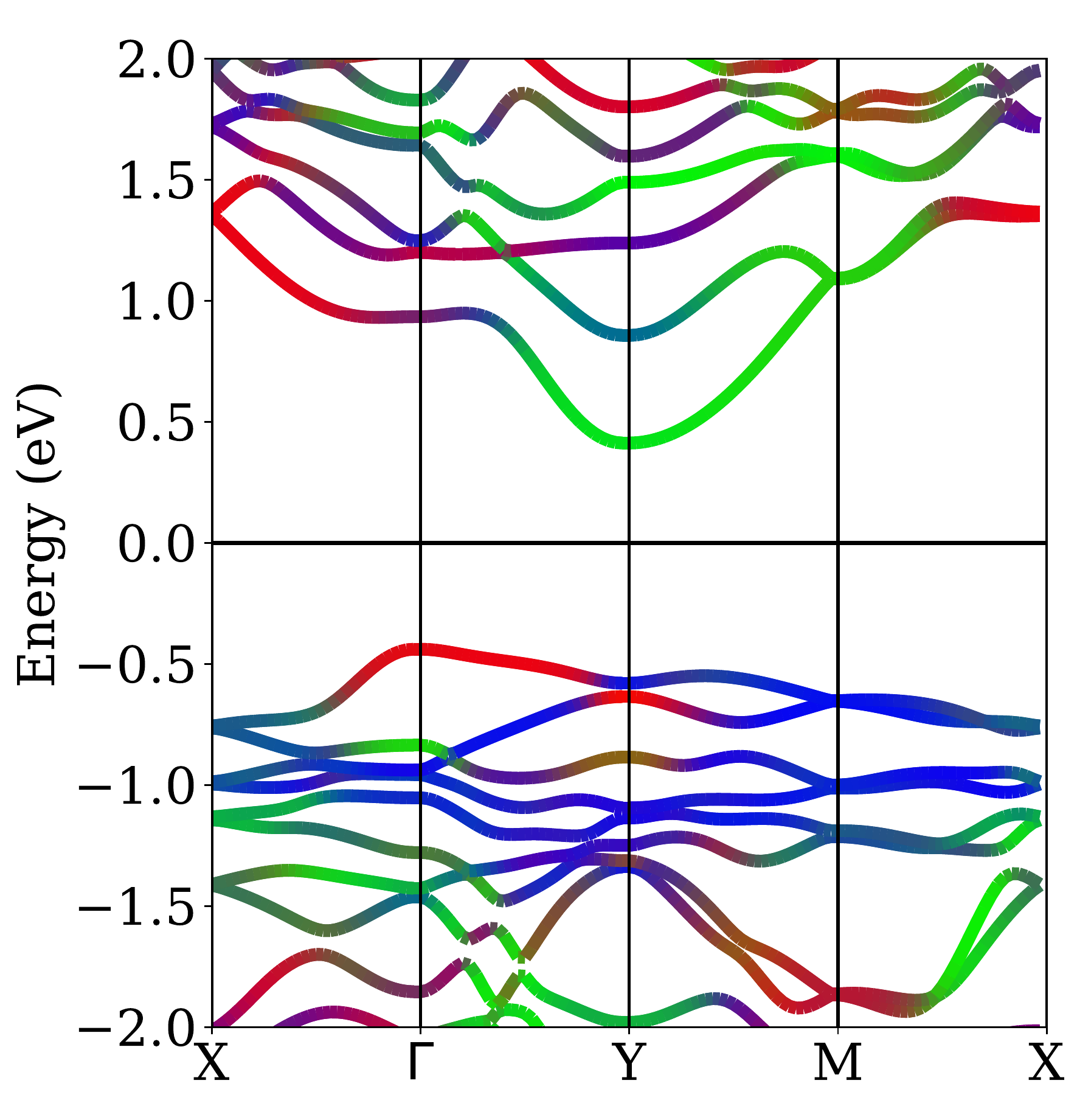}}\\
\vspace{0.3cm}
\sidesubfloat[]{\includegraphics[width=4cm,scale=1.0,clip, keepaspectratio]{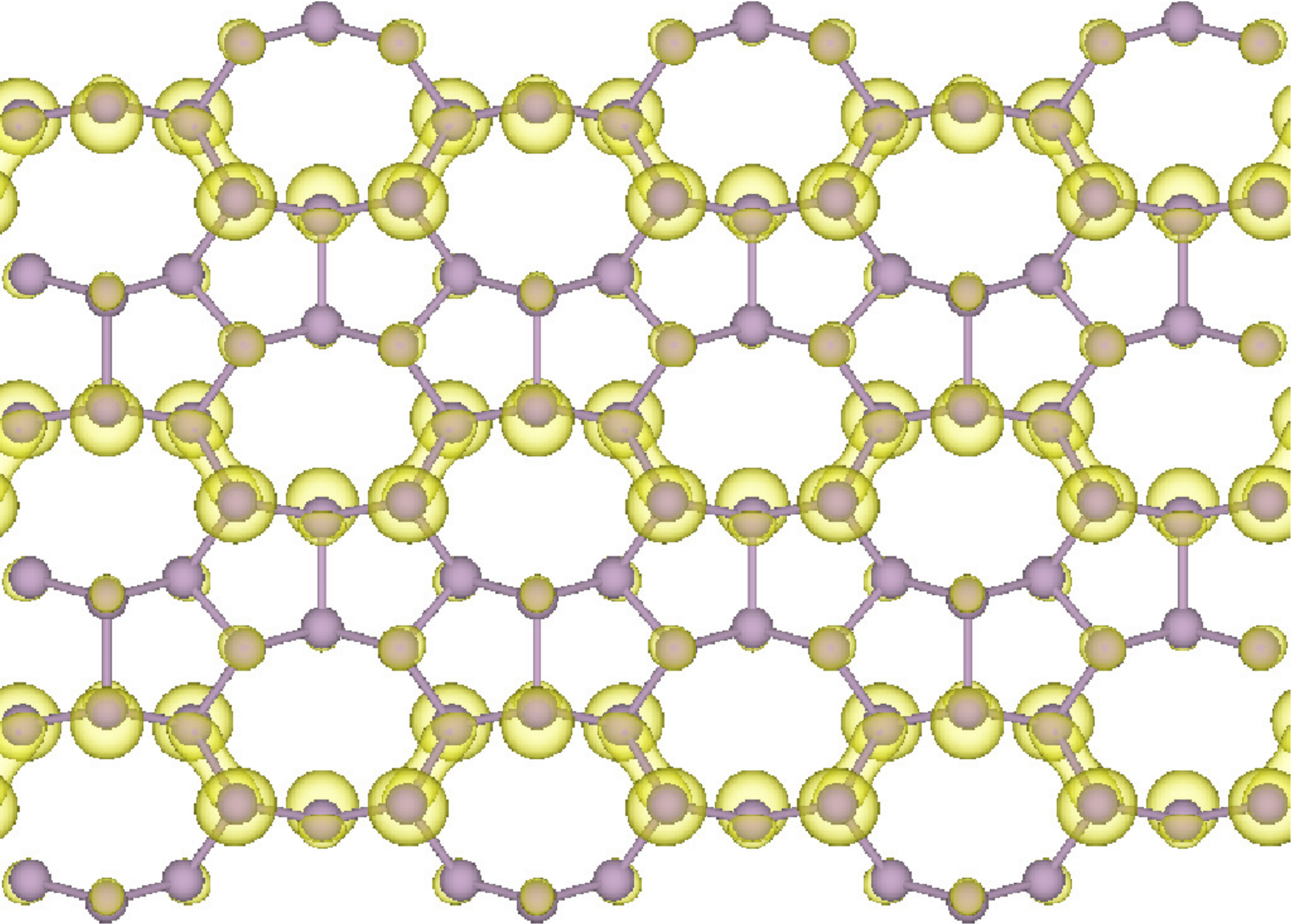}}\hfill
\sidesubfloat[]{\includegraphics[width=4cm,scale=1.0,clip, keepaspectratio]{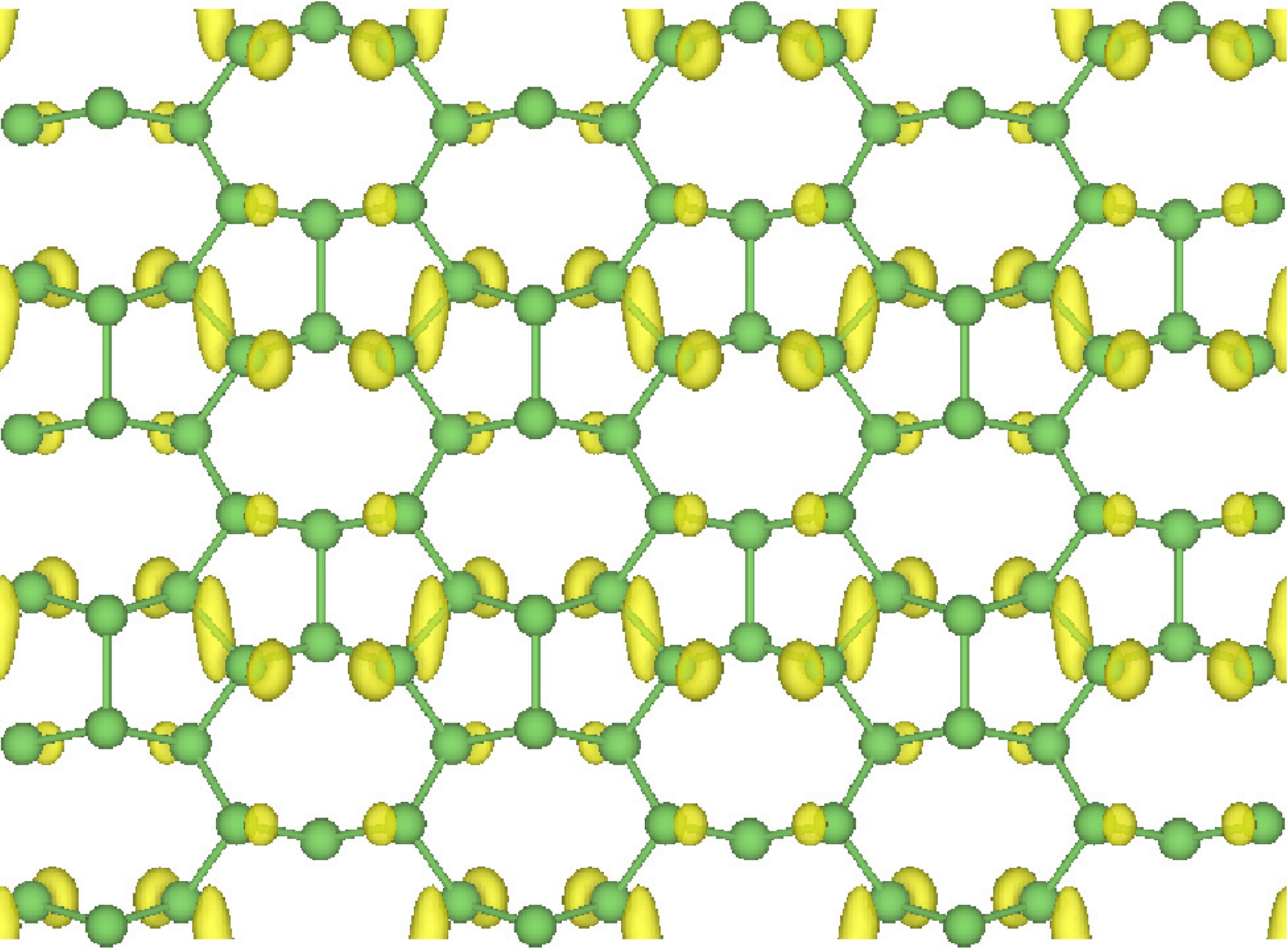}}\hfill
\sidesubfloat[]{\includegraphics[width=4cm,scale=1.0,clip, keepaspectratio]{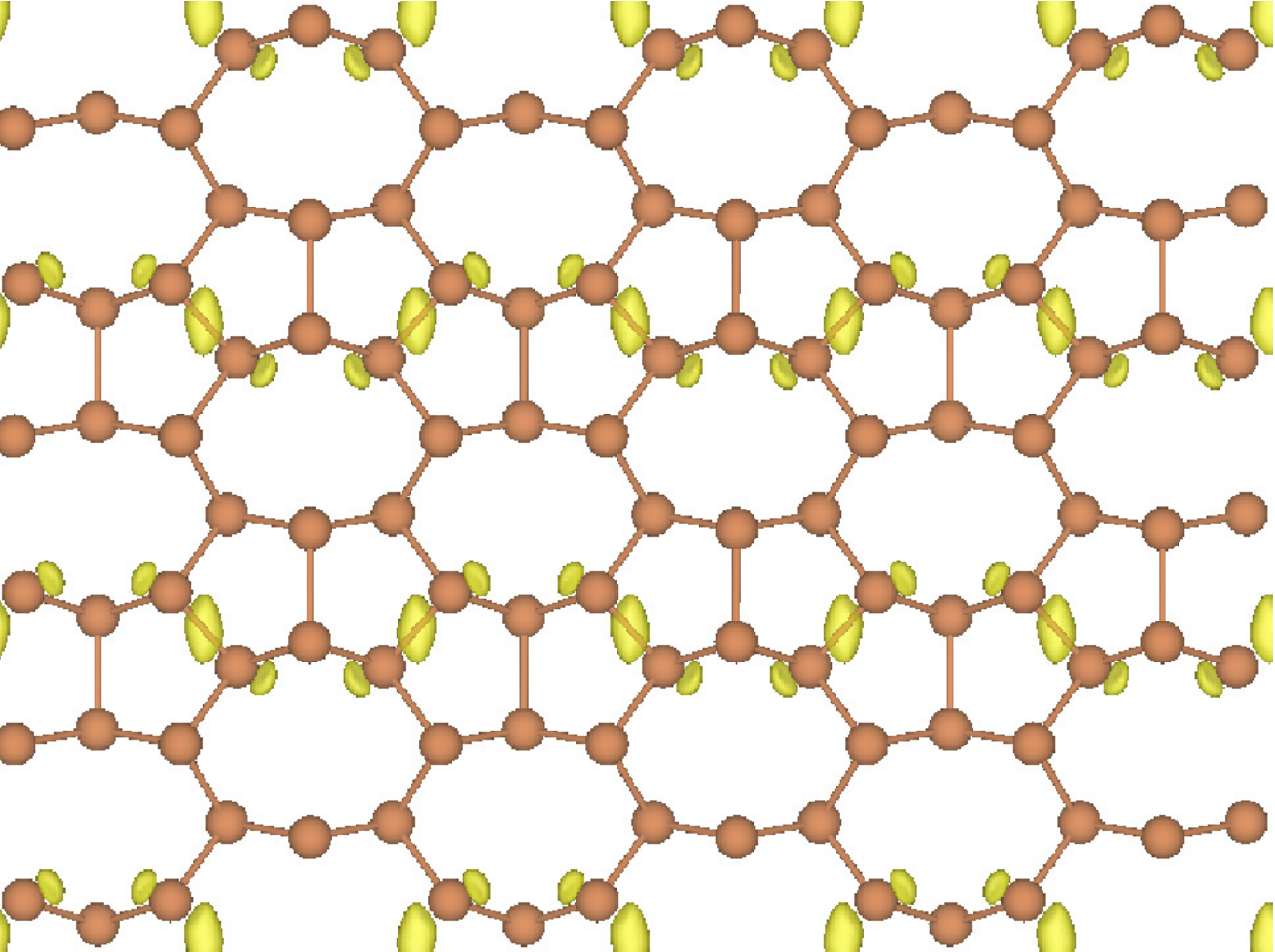}}
\caption{\label{fig:bandas-sem-tensao}
Electronic band structure, with different orbital contributions of hexagonal (a) P, (b) As, and (c) Sb and pentaoctite (d) P, (e) As and (f) Sb at zero strain.  Red, green, and blue colors represent contribution from p$_x$, p$_y$, and p$_z$ orbitals, respectively. Band decomposed charge density at VBM for pentaoctite phases of (a) P, (b) As and (c) Sb at zero strain. Isosurface values are 5x10$^{-4}$ e/${\AA}^3$} 
\end{figure*}

In hexagonal Sb, VBM is at $\Gamma$-point and CBM between $\Gamma$ and
M points as seen in Fig. \,\ref{fig:bandas_hexagonal} (c).  Hexagonal
antimonene has a band gap of 1.02 in good agreement with other
calculations\,\cite{AngChem2016,JMCC2016}. Finally in
pentaoctite-Sb\,\cite{Ciraci2019} the energy fundamental band gap is
0.83 eV with GGA and 1.01 eV with HSE as seen in Table
\,\ref{table:parameters}. As a general conclusion, the band gaps in
pentaoctite phases are larger than in hexagonal phases for P, As and
Sb. This difference in the band structure and band character can be
understood considering the different buckling and hybridization.

Fig.\ref{fig:bandas-sem-tensao} the band decomposed charge density at VBM for
pentaoctite phases of P, As and Sb at zero strain. One can see that
for P, Fig.\ref{fig:bandas-sem-tensao} (d) the main contribution comes from
orbitals p$_z$. For As, as shown in Fig.\ref{fig:bandas-sem-tensao} (e), the main
contribution comes from orbitals p$_x$. Finally, for Sb the main
contribution comes from orbitals p$_x$ as seen in
Fig.\,\ref{fig:bandas-sem-tensao} (f). The flatening of the band structure is due
to the different hybridization within the octagon-pentagon rings, as
discussed in Ref.\,\cite{LimaJPCM2019,LimaNL2016}.

\begin{figure}[ht!]
\begin{center}
  \includegraphics[width=7cm, clip, keepaspectratio]{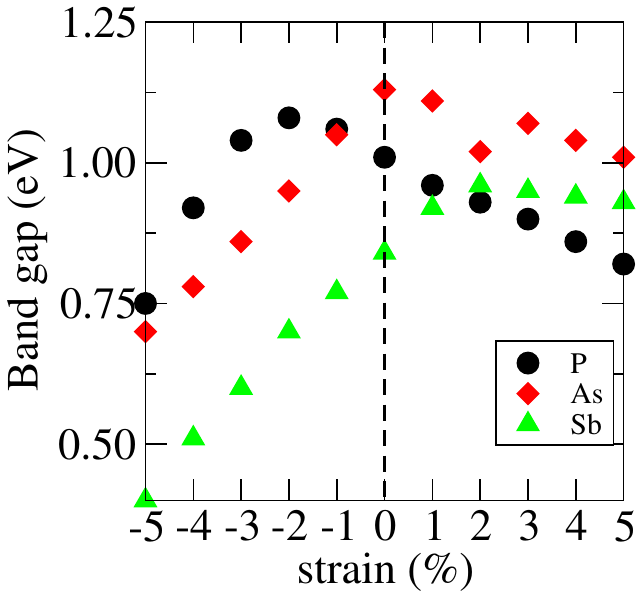}
  \end{center}
 \caption{\label{fig:strain} Electronic band gap as a function of strain for phosphorene, arsenene and antimonene. Positive (negative) values mean tensile (compressive) strain.}
\end{figure}

It is known that external strain can be used to tune and modify the
electronic band structure of materials. As all investigated structures
show indirect band in the pentaoctite phase. Therefore we have applied compressive and tensile
strains between -5\% and +5\%. Fig.\,\ref{fig:strain} shows that indeed the band gap is
sentisive to strain. However, all structure are more sensitive to
negative (compressive) strain than to positive (tensile) strain. In
particular, pentaoctite-As shows the largest variation for tensile
strain.

\begin{figure*}[ht!]
\centering
\sidesubfloat[]{\includegraphics[width=5cm,scale=1.0,clip,keepaspectratio]{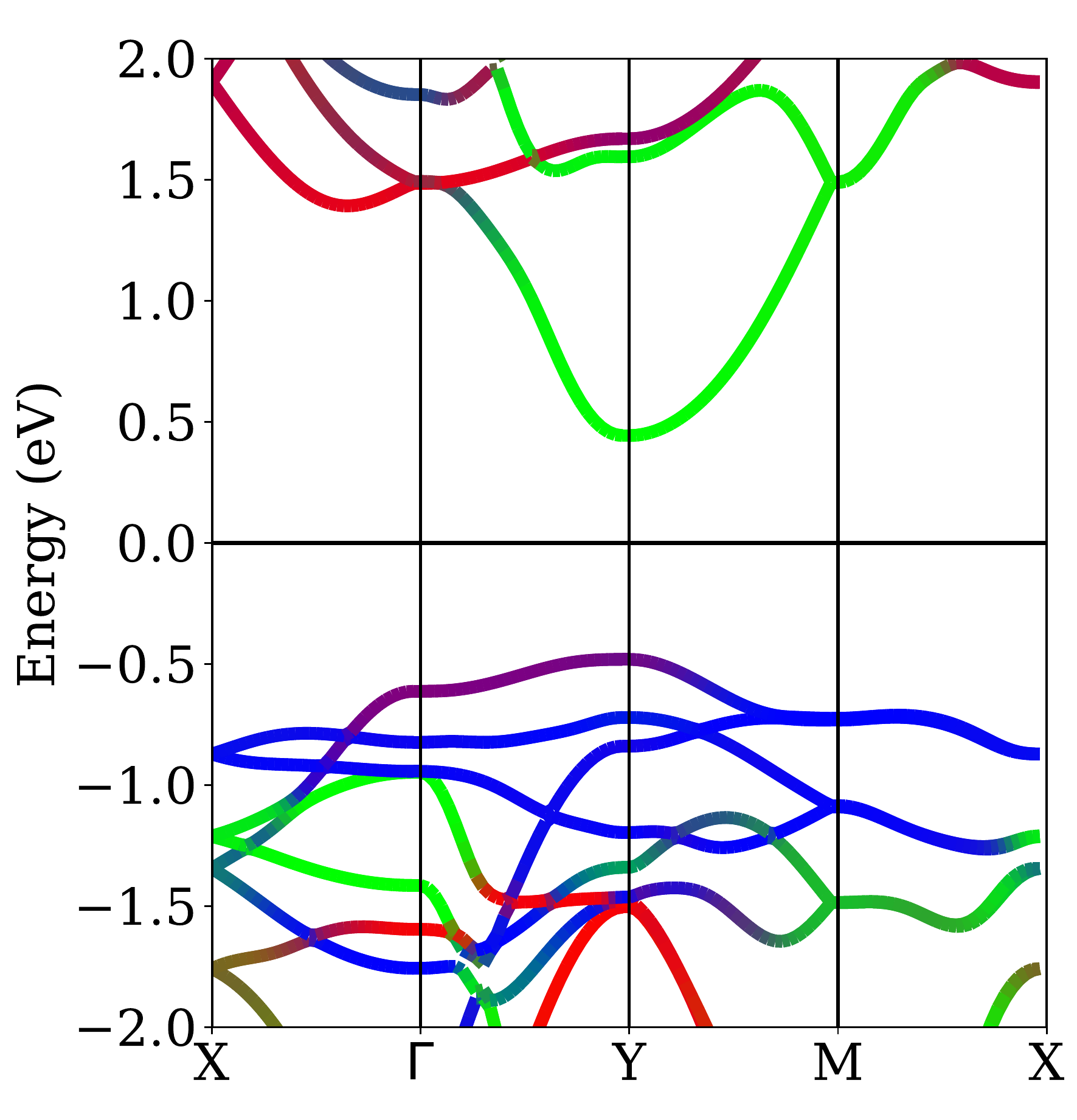}}
\sidesubfloat[]{\includegraphics[width=5cm,scale=1.0,clip,keepaspectratio]{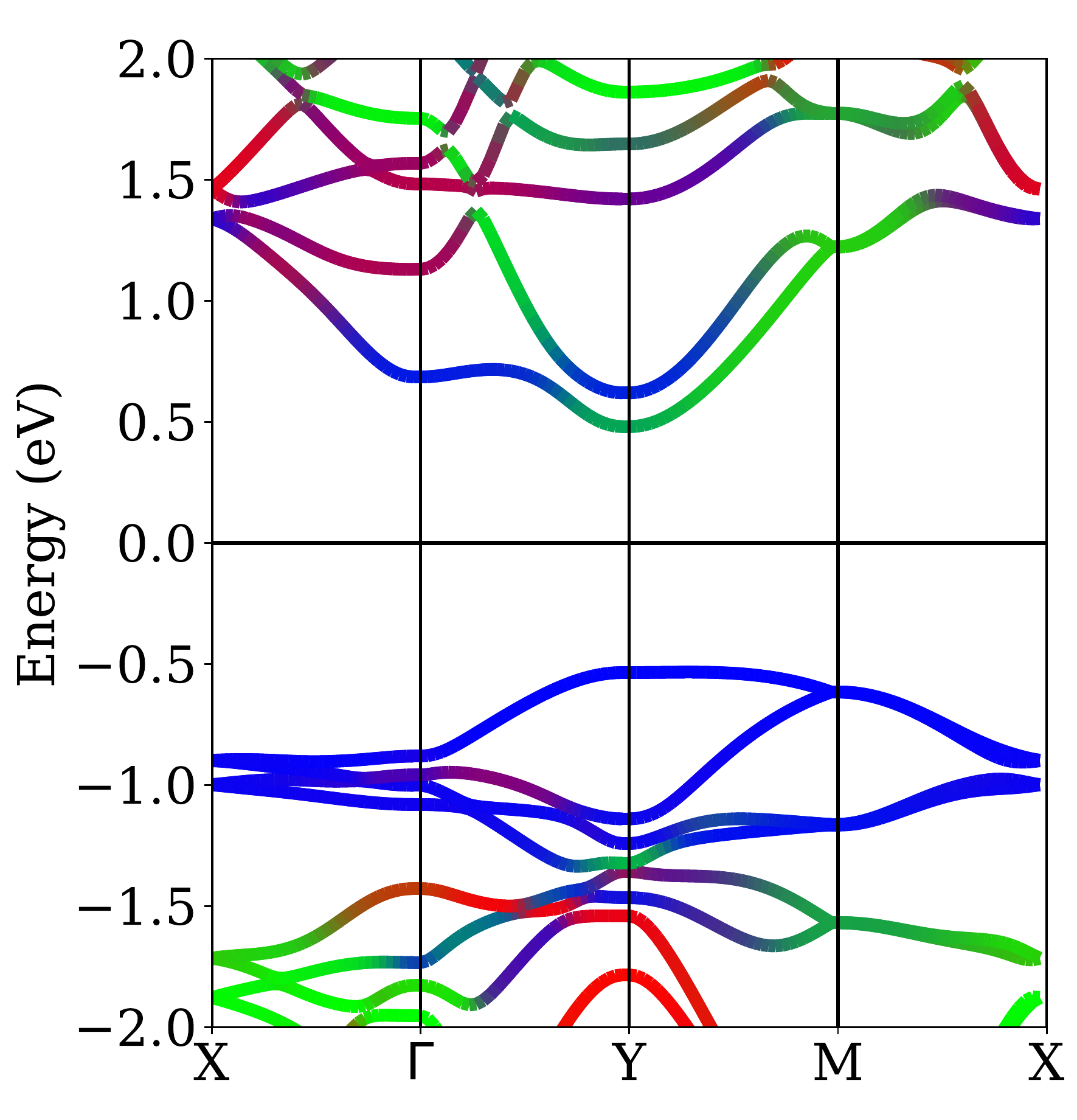}}
\sidesubfloat[]{\includegraphics[width=5cm,scale=1.0,clip,keepaspectratio]{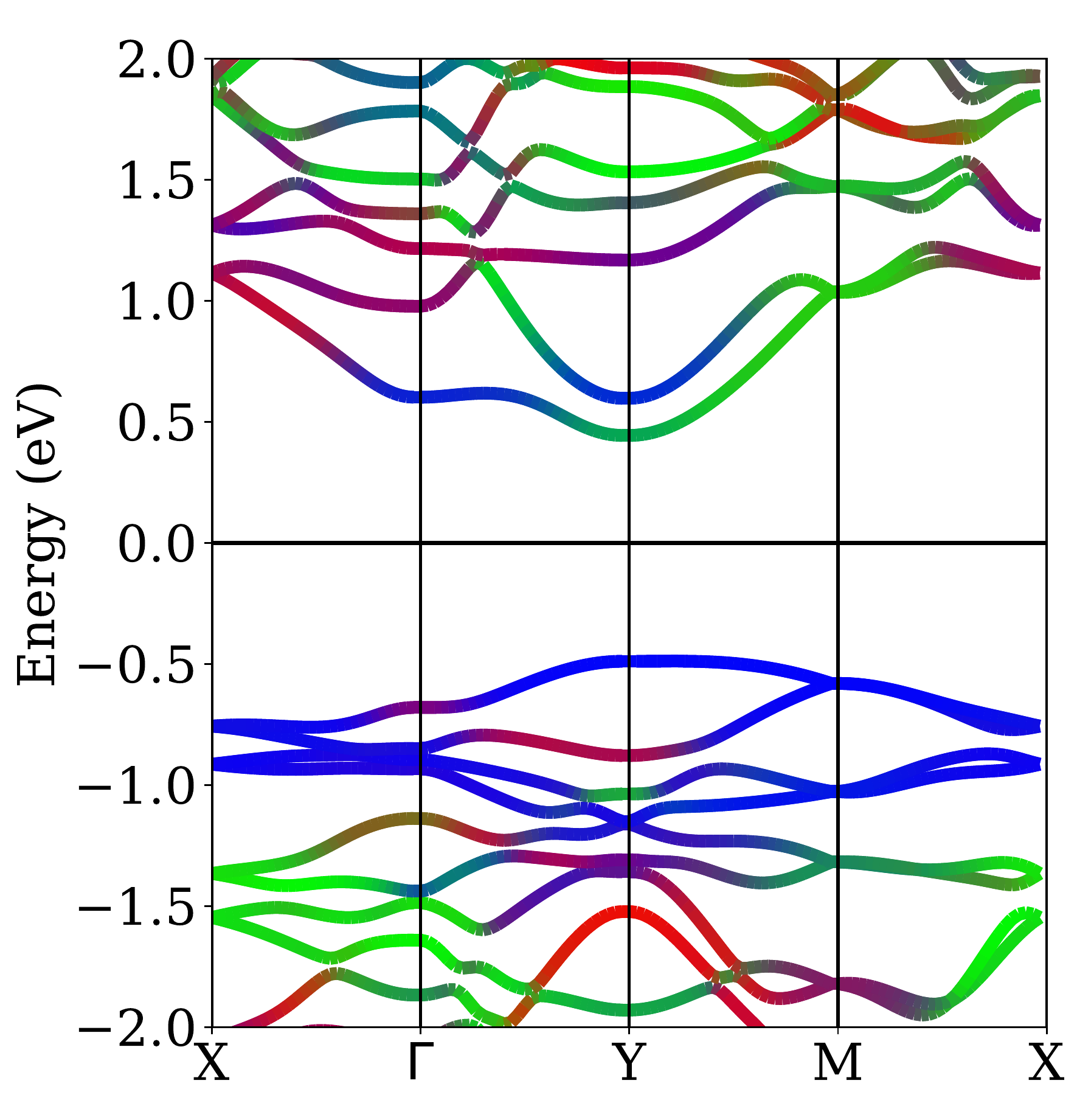}}
\caption{\label{fig:bandas-com-tensao} Band structure of pentaoctite phase (a) P, (b) As, and (c) Sb, under -4\%, 
5\%,and 5\% strain condition, respectively. Red, green, and blue colors represent contribution from p$_x$, p$_y$, and p$_z$ orbitals, respectively.} 
\end{figure*}

The resulting band structures of strained pentaoctites are shown in
Fig.\,\ref{fig:bandas-com-tensao}.  The specific strain show the
transition from indirect to direct bandgap with a VBM and CBM at the Y
point. In pentaoctite-P compressive strains $>$4\%
(Figs.\,\ref{fig:bandas-com-tensao} (a)) leads to a direct
bandgap, whereas tensile strains $>$5\% transform pentaoctite-As and
pentaoctite-Sb (Figs.\,\ref{fig:bandas-com-tensao} (b) and (c))
into a direct bandgap semiconductor.

\begin{figure}[ht]
  \begin{center}
  \begin{tabular}{cc}
\includegraphics[width=4cm,clip, keepaspectratio]{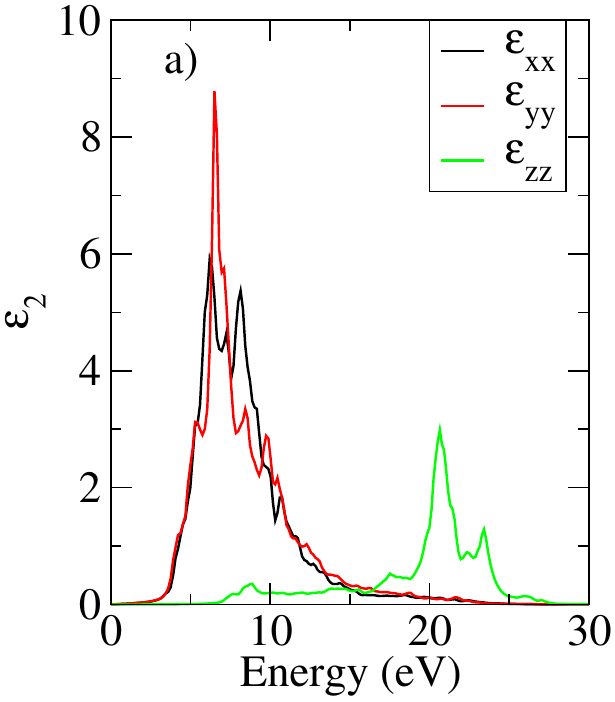} &
\includegraphics[width=4cm,clip, keepaspectratio]{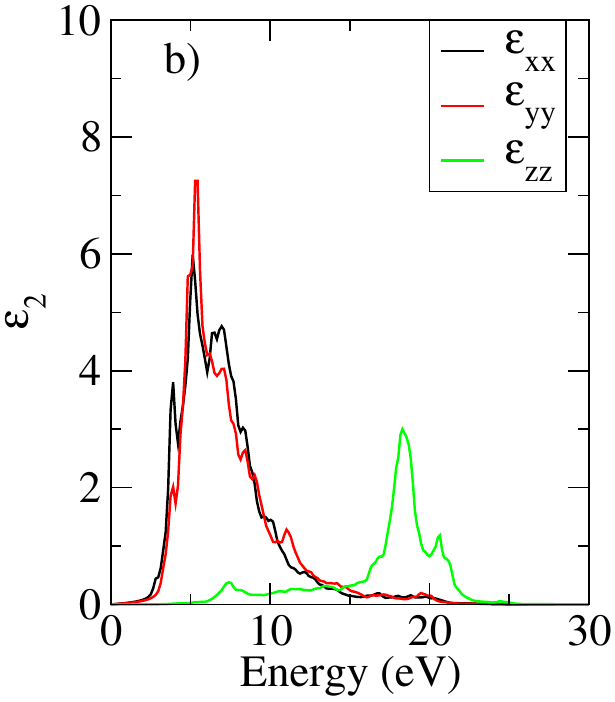}\\
\includegraphics[width=4cm,clip, keepaspectratio]{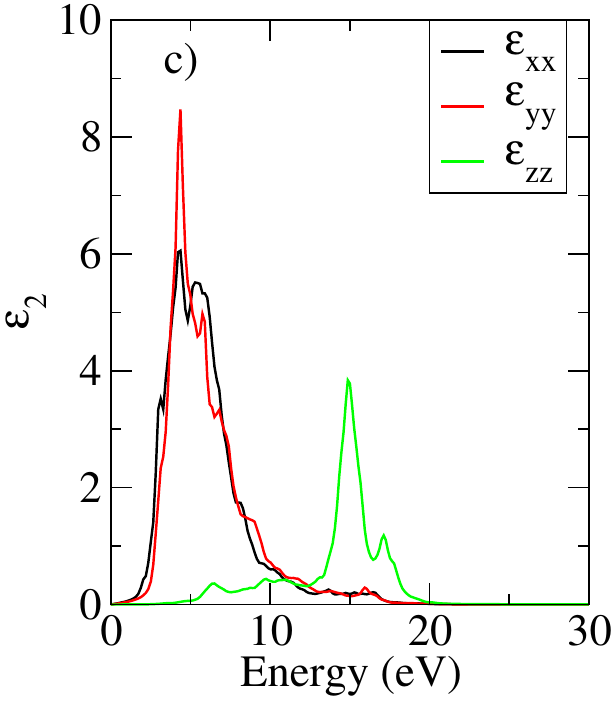}&
\includegraphics[width=4cm,clip, keepaspectratio]{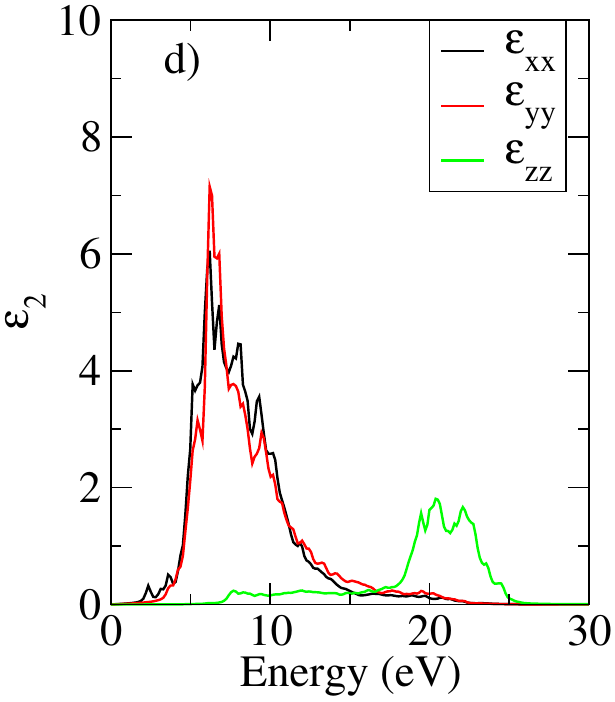}\\
\end{tabular}
  \end{center}
  \caption{\label{fig:diel_GW} Imaginary part of the dielectric function within GW$_0$ approximation for pentaoctite a) phosphorene, b) arsenene, c) antimonene, d) phosphorene under compressive strain of  $>$4\%.
  Light propagation parallel to the pentaoctite layer 
  is denoted as $\epsilon_{xx}$ and $\epsilon_{yy})$. Light propagation perpendicular to 
  the pentaoctite layer is denoted as $\epsilon_{\rm perp}=\epsilon_{zz}$.}
  \end{figure}

It has been predicted that several two-dimensional materials behave
like topological insulators, which implies the presence of conducting
edge states carrying two counter-propagating spin-polarized
currents\,\cite{Z2}. Due to constraints of time-reversal
symmetry, these conducting edge states are protected against
backscattering, making them suitable for spintronics
applications. Although graphene was initially proposed behave like a
topological insulator, the spin-orbit coupling (SOC) in this material
is very weak. In group V-materials, hexagonal and pentaoctite bismuth
monolayers have been predicted to be a 2D topological insulator
(TI)\,\cite{Kou:NL,Rivelino2015,AsiaNat,Reis2017,JPCC2020,LimaJPCM2019,LimaNL2016}.
In order to investigate whether the pentaoctite sructures behave like
topological insulators we have calculated the Z$_2$ invariant
according to Ref.\,\cite{Z2}. On the contrary of pentaoctite
bismuthene, our investigated structures are not topological
insulators.

Finally, the dielectric function of pentaoctite layers were
investigated by calculating the imaginary part of the dielectric
function at G$W_0$ level. This means that the Green's functions are
calculated iteratively whereas the Coulomb potential W is kept at the
DFT level. The imaginary part of the dielectric function is calculated
directly from the electronic structure through the joint density of
states and the momentum matrix elements occupied and unoccupied
eigenstates according Ref.\cite{Shishkin:07}.

We show the dielectric function calculated within the G$W_0$
approximation in Fig.\,\ref{fig:diel_GW}. The parallel
$\varepsilon_{\rm xx}$ and $\varepsilon_{\rm yy})$ and perpendicular
$\varepsilon_{zz}$ components of the the imaginary part of the
dielectric function $\varepsilon_2$ are shown. The parallel
components corresponds to the propagation of the external
electromagnetic field parallel to the pentaoctite plane while
$\varepsilon_{zz}$ corresponds to the field perpendicular to the
plane. Because of optical selection rules, anisotropy in the optical
spectra is seen. Anisotropy has also been reported in layered
monochalcogenide of germanium sulfide (GeS) \cite{GeS}, black
phosphorous\,\cite{blackP} and bismuthene\cite{Ciraci2019,JPCC2020}
and germanene\,\cite{PCCP2020}. The systems with a gap show finite
absorption limits for both parallel and perpendicular directions with
larger intensity for the $(\varepsilon_{\|}$ component.

In the imaginary part of the dielectric function, the energy onsets
are 2.6 as seen in Fig.\,\ref{fig:diel_GW}(a), 2.1 seen in
Fig.\,\ref{fig:diel_GW}(b) and 1.51 eV seen in
Fig.\,\ref{fig:diel_GW}(a) for P, As and Sb,
respectively. Additionally, there are only one main peak around 7.0,
5.27 and 6.2 eV for P, As and Sb, respectively. Materials with band
gaps below the 1.65 eV absorb well in the infrared (IR) region of the
spectrum. Therefore, antimonene will start to absorb electromagnetic
radiation in the IR region as an optical material, but absorption in
antimonene will be strongest in the UV part of electromagnetic
spectrum. In Fig.\,\ref{fig:diel_GW}(d) we show the dielectric
function for strained antimonene. The strain shifts the spectrum to
smaller energies, as expected.

\section{Conclusions}

We have performed density-functional theory calculations for group V
allloropes.  We show that pentactotie structures of these materials
are stable. The electronic properties show a sizeable band gap with
absorption spectrum in the visible region. We suggested that such
defects can be useful as building blocks for group-V electronics.

\section{Acknowledgements}

We acknowledge the financial support from the Brazilian Agency CNPq
and FAPEG (PRONEX 201710267000503) and German Science Foundation (DFG)
under the program FOR1616. The calculations have been performed using
the computational facilities of Supercomputer Santos Dumont and at QM3
cluster at the Bremen Center for Computational Materials Science and
CENAPAD.

%\bibliography{references}

\end{document}